\documentclass[11pt]{article}

\usepackage{booktabs} 
\usepackage{amsmath}
\usepackage{hyperref}
\usepackage{graphicx}
\usepackage{multirow}
\usepackage{subcaption}
\usepackage{bchart}
\usepackage{listings}
\usepackage{color}
\usepackage{comment}
\usepackage{times}
\definecolor{dkgreen}{rgb}{0,0.6,0}
\definecolor{gray}{rgb}{0.5,0.5,0.5}
\definecolor{mauve}{rgb}{0.58,0,0.82}

\usepackage{coling2020}

\usepackage{array}

\usepackage{enumitem}

\usepackage[ruled,vlined]{algorithm2e}

\begin{document}

\title{Audio Adversarial Examples: Attacks Using Vocal Masks} 

\author{Tay Kai Yuan, Ng Hui Xian Lynnette, Chua Wei Han, \\ \textbf{Lucerne Loke, Ye Danqi, Chua Wan Jun Melissa} }

\maketitle

\begin{abstract}
We construct audio adversarial examples on automatic Speech-To-Text systems . Given any audio waveform, we produce an another by overlaying an audio vocal mask generated from the original audio. We apply our audio adversarial attack to five SOTA STT systems: DeepSpeech, Julius, Kaldi, wav2letter@anywhere and CMUSphinx. In addition, we engaged human annotators to transcribe the adversarial audio. Our experiments show that these adversarial examples fool State-Of-The-Art Speech-To-Text systems, yet humans are able to consistently pick out the speech. The feasibility of this attack introduces a new domain to study machine and human perception of speech.

\end{abstract}

\section{Introduction}
With the advent of virtual assistants like Google Assistant, Apple's Siri and Amazon's Alexa, more attention has been brought to the space of Speech-To-Text (STT), where natural language commands are converted into computer texts. These texts carry information from the natural language command, authenticating the virtual assistant to execute an action, such as delivering news, even order products from an online store \cite{voicecommerce}.

Adversarial machine learning is a sub-field of machine learning that has gained much attention in recent years. In adversarial machine learning, malicious inputs exploit weaknesses in a trained model or training regime to produce undesired and unforeseen behaviors. Goodfellow et. al. \cite{goodfellow2014explaining} proposed the Fast Gradient Sign Method for generating adversarial inputs on image classifiers. Jang et. al. \cite{jang2017objective} further perfected gradient based techniques by proposing a technique that could generate minimal perturbations on images in relation to its features, demonstrating the ease of exploitation of image classifiers. 

Most of the research of adversarial machine learning has been centered around the domain of computer vision, and very few techniques are available to create adversarial examples in STT. Carlini and Wagner \cite{Carlini2018AudioAE} proposed a gradient based technique adopted from image classification domain and applied it on audio samples, creating adversarial examples by iteratively minimising perturbations introduced into the audio waveform, while ensuring that the waveforms are transcribed as another message. Schoenherr et al. \cite{Schoenherr2019} uses of psychoacoustic hearing properties to generate perturbations to the audio, to create audio below the hearing threshold such that the attack is inaudible to the listener. The generated audio renders the transcription through Kaldi ineffective. Zhang et al. \cite{dolphin} introduced the dolphin attack that modulates voice on ultrasonic carriers to achieve inaudible attack vectors. While it is validated on popular speech recognition signals, this attack is costly to construct.

\subsection{Perception of Speech}
Just like how our brains fill in the gaps in our vision, our brains formulate speech to allow us to understand the speaker better \cite{auditoryillusion}. In a noisy environment, if we pay enough attention to a speaker and his speech, it is relatively easy for our brains to follow the words. Following this concept, as long as any background noise is below a certain threshold in decibels, humans should not fail to formulate a speaker's speech\cite{Darwin2007ListeningTS}. On the other hand, STT systems often rely largely on several quantitative qualities of speech in order to accurately transcribe speech. Lacking the language ability to fill in words that were not heard clearly in an audio, STT systems may fail to transcribe accurately in noisy situations.

Human speech sounds are produced by the different shapes of the vocal tract. Mel-Frequency Cepstral Coefficients (MFCCs) \cite{MFCC} and vocal masks are frequently used to represent the shape of the vocal tracts, which manifests itself in the sort time power spectrum, generated by performing Short-Time Fourier Transform (STFT) on the audio signal. STT systems learn these quantitative representation of words on these audio waveforms to transcribe subsequent audio files. Recent work on convolutional neural networks on mel-frequency spectrograms have shown remarkable accuracy in voice separation \cite{simpson2015deep}\cite{Ikemiya_2016}. This is largely due to extensive learning of features of vocal masks from Mel-Frequency spectrograms\cite{Lin2018SingingVS}, and how they differ from other noise, such as background music. However, these properties allow adversaries to employ specific attacks against these deep learning systems.

\subsection{Contributions}
In this paper, we propose a novel method of creating adversarial examples on audio signals that attacks five State-Of-The-Art (SOTA) STT systems. These adversarial examples are transcribed as a different message by Speech-To-Text systems, while humans are able to decipher the speech in the signal, rendering STT systems inadequate. We are able to achieve this by overlaying a vocal mask on top of the original audio, making use of the inability of these neural networks to differentiate a vocal mask from the original speech, resulting in an average Word Error Rate (WER) of 0.64. In comparison, the same adversarial audio can be transcribed by human annotators with an average WER of 0.28. This is an end-to-end attack, that operates directly on the raw samples that are used as inputs to the neural networks.

The tasks that this paper attempts are: 
\begin{enumerate}
    \item Generation of adversarial audio examples at different decibel levels using Mel-Frequency Cepstral Coefficients (MFCC) properties 
    \item Generation of targeted adversarial audio examples using the Carlini-Wagner audio attack \cite{Carlini2018AudioAE}
    \item Transcription of adversarial audio examples using five SOTA speech-to-text transcription neural networks.
    \item Comparing transcription output from neural networks with human transcription
\end{enumerate}

The State-Of-The-Art Speech-To-Text Systems attacked in this paper are: 
\begin{enumerate}
\label{sec:sttsystem}
    \item DeepSpeech\cite{Hannun2014DeepSS}. An end-to-end speech recognition system built upon Baidu's DeepSpeech architecture. To enhance its capabilities, DeepSpeech learns a function that is robust to background noise, reverberation and speaker variation.
    \item Kaldi\cite{Povey_ASRU2011}.  One of the oldest toolkit for Automatic Speech Recognition (ASR), which integrates with Finite State Transducers and provide extensive linear algebra support built upon the OpenBLAS library. Kaldi supports Hidden Markov Models, Gaussian Mixture Models, and neural-network based acoustic modelling.
    \item Wav2letter@anywhere\cite{wav2letter@anywhere}. Uses ArrayFire tensor library and flashlight machine learning library, enabling training of the LibriSpeech dataset to be completed within minutes. The language model formation uses Time-Depth Separable Convolutions, while the decoder module is a beam-search decoder.
    \item Julius\cite{julius}. A large vocabulary continuous speech recognition decoder software based on n-gram and context-dependent hidden markov models developed by the Japanese since 1997\cite{jilushmm}. The decoding algorithm is based on a two-pass tree-trellis search that incorporates major decoding techniques including tree lexicons, n-gram factoring, enveloped beam search, and deep neural networks.
    \item CMUSphinx\cite{sphinx}. A collection of speech recognition tools aiming to overcome the constraints of isolated words and small vocabularies. Its acoustic model is based on Hidden Markov Models, and speech constraints overcome through function-word-dependent phone models and generalised triphone models\cite{sphnixkaifu}.
\end{enumerate}

To facilitate future work, we make our dataset available. We encourage the reader to listen to our audio adversarial examples\footnote{https://drive.google.com/open?id=1ixwuvPrk1H-hveX5HNSWWYGSzx-lWYnC}.

\section{Methodology}
\label{sec:methodology}
This transcription performance comparison of audio adversarial examples between SOTA neural network-based STT models and human annotators is empirically evaluated with a series of experiments. This section describes the generation of audio adversarial examples and defines the threat model and evaluation metric, which will subsequently be used for performance comparison between the produced transcription and the original transcription.

\subsection{Threat Model}
Given an audio waveform $x$, target transcription $y$, and SOTA STT model $C$, and the human transcription $D$, our task is to construct another audio waveform $x' = x + \delta$ so that $D(x') = y$ but $C(x') \neq y$. 

\subsection{Distortion Metric}
To quantify the distortion introduced by an adversarial audio, we use the Decibels (dB) measure. dB is a logarithmic scale that measures the relative loudness of an audio sample. This measure is a relative measure, hence we use the original waveform $x$ as the reference point for the adversarial audio $x_1$.

$$dB(x_1) = 10 \log \frac{S(x_1)}{S(x)},$$
where $S$ is the function that maps $x$ to the intensity of the soundwaves.

In this paper, most of the distortion is quieter than the original signal, the distortion is thus a negative number. 

This metric may not be a perfect measure of distortion, however the audio distortion may be imperceptible to the humans, as we will show in our experiments, where human annotators fare better than SOTA STT systems. 

\subsubsection{Accuracy Metric}
In calculation of the accuracy of the transcription, we calculate the Word Error Rate (WER) of the produced transcription against the original transcription provided by the TIMIT dataset.

The WER is a common measure of accuracy in performance measurement in STT systems. It is derived from the Levenshtein distance, which measures the difference between a sequence of characters through the minimum number of edits. By measuring the difference in words, transcribed word sequence can have different lengths from the ground truth. However, this measure does not detail the nature of transcription errors. In our experiments, we calculate the Levenshtein distance using the Wagner-Fisher algorithm \cite{10.1145/321796.321811}. It should be noted that the lower the WER, the better the transcription.
$$WER = \frac{S + D + I}{N},$$
where $S$ is the number of substitutions,
$D$ is the number of deletions,
$I$ is the number of insertions, and
$N$ is the number of words in the reference.

\subsection{Dataset}
We used the DARPA-TIMIT Acoustic-Phonetic Continuous Speech Corpus \cite{timit} for construction of adversarial audio and experiments. This dataset contains time-aligned transcriptions of phonetically rich spoken American English sentences. A hand-verified transcription of the corpus is made available, which serves during the calculation of the model accuracy. This dataset is suitable for our studies as we are studying the spoken speech and our human annotators are well-versed in the English language.

\subsubsection{Generation of obfuscated audio}
\label{sec:obfusgeneration}

We generate obfuscated audio with the python library \textit{Pydub} \cite{pydub}. \textit{Pydub's AudioSegment} object provides several methods that allow easy manipulation of audio, including reading and writing audio files, as well as reversing audio files and overlaying them on other audio files.

Let $x$ be the original audio waveform, and $\delta$ be the adversarial audio signal generated using \textit{AudioSegment} as an attack to the original waveform. We generate $\delta$ by reversing $x$. We then overlay the audio $\delta$ on to $x$, that is, $x' = x + \delta$.

As decibels is a measure of the intensity of the sound, $dB(\delta) = dB(x)$. We generate five adversarial audio waveforms by varying the intensity of the sound, each time decreasing the intensity of $\delta$. The waveforms shall be represented as: $dB(\delta_p) = dB(\delta) - p$, where $p$ is the amount of decibels that $\delta_p$ differs from $\delta$. For example, if the adversarial audio was decreased by 5 decibels, $dB(\delta_{-5})= dB(\delta) - 5$.

\subsubsection{Model set up}
We constructed the experimental set up of the neural network models as follows:
\begin{enumerate}
    \item \textit{DeepSpeech}: Used the provided pre-trained English model, an n-gram language model trained from a corpus of 220 million phrases with a vocabulary of 495,000 words. This was trained on the LibriSpeech\cite{librispeech}, Fisher\cite{Cieri2004TheFC}, Switchboard\cite{225858} and Common Voice English\cite{ardila2019common} datasets, for 233,784 steps where the best validation loss were selected at the end of 75 epochs.
    \item \textit{Julius}: Used the provided pre-trained n-gram language model, in a hybrid Deep Neural Network with Hidden Markov Model based architecture (LM+DNN-HMM). The model with a 262,000 word dictionary and 32 bit Language Model\cite{L_scher_2019}\cite{Lee2009RecentDO}.
    \item \textit{Kaldi}: Used the Kaldi pre-trained Aspire Chain Model\cite{aspire} with already compiled HCLG sequence of phoneme decoding graph for inference, trained on Fisher English\cite{Cieri2004TheFC}. HCLG is a hidden-markov finite state transducer representing the lexicon, grammar and phonetic contexts.
    \item \textit{Wav2letter@anywhere}: Used the provided inference platform that pre-trains BERT\cite{devlin2018bert} models with Librispeech dataset\cite{librispeech}.
    \item \textit{CMUSphinx}: Used the provided inference platform on python bindings PocketSphinx \cite{sphinx}, with the pre-trained US English constructed from Wall Street Journal data using hidden-markov finite state transducers.
\end{enumerate}{}

\subsubsection{Baseline Adversarial Attack}
\label{sec:cwaudioattacks}
We used the Carlini-Wagner (CW) audio adversarial attack as the baseline \cite{Carlini2018AudioAE} audio adversarial attack. This attack adds a small perturbation that are quieter than the original signal to the original audio which changes the transcribed result when passed through DeepSpeech. We show that our audio attacks surpass the CW attack.

\subsubsection{Human Audio Transcription}
We engaged seven people from ages 25-30 to transcribe the audio for us. Each person transcribed 34 audio files of a single type of obfuscation. These audio files were randomly chosen from 2 dialect groups of the DARPA-TIMIT corpus' test directory. As the audio files were primarily in English, to reduce problems with understanding the text, we engaged people whose first spoken and written language is English. We asked the annotators to listen to the samples via headphones. The task was to type all words of the audio sample into a blank text field without assistance from auto-complete or grammar and spell-checking. Annotators often repeated the audio samples in order to enter a complete sentence. In a post-processing phase, we removed symbols and new lines from the transcript before calculating the WER.

\subsection{Results and Analysis}
We present the mean WER for the various transcription systems in Table \ref{tab:werresults}. 

We note that DeepSpeech has a WER that is greater than 1.0 because there are more words than the number of words in the reference speech. This is likely due to the fact that DeepSpeech was trained on a dataset that has more words than the TIMIT dataset. In terms of audio adversarial attacks, we note that the attack of $x + \delta_{-15}$ produces the best balance between fooling neural STT systems and human identification of the audio. At the same time, $x + \delta{-15}$ has a higher WER than $x + \text{CW}$, proving our attack to be more effective at fooling neural network systems as compared to the baseline Carlini-Wagner attack.

\begin{table*}[ht]
\centering
\begin{tabular}{p{2cm} p{1.5cm} p{1.5cm} p{1.5cm} p{2cm} p{1.5cm} p{1.5cm}}
  \hline
  Audio Files & DeepSpeech & Julius & Kaldi & wav2letter@ \newline anywhere & CMUSphinx & Humans\\  
  \hline
  $x$                   & 0.13 (0.15) & 0.76 (0.31) & 0.32 (0.20) & 0.16 (0.17) & 0.35 (0.31) & 0.07 (0.10)  \\
  $x + \delta_0$        & 1.03 (0.26) & 0.99 (0.19) & 0.90 (0.18) & 0.87 (0.22) & 1.26 (0.40) & 0.77 (0.36)  \\
  $x + \delta_{-5}$     & 0.82 (0.27) & 0.93 (0.17) & 0.77 (0.22) & 0.65 (0.22) & 1.02 (0.40) & 0.36 (0.30)  \\
  $x + \delta_{-10}$    & 0.63 (0.35) & 0.85 (0.26) & 0.62 (0.27) & 0.42 (0.23) & 0.94 (0.41) & 0.28 (0.29)  \\
  $x + \delta_{-15}$    & 0.35 (0.28) & 0.78 (0.29) & 0.48 (0.28) & 0.26 (0.20) & 0.68 (0.32) & 0.10 (0.11)  \\
  $x + \delta_{-20}$    & 0.23 (0.24) & 0.79 (0.29) & 0.37 (0.23) & 0.18 (0.18) & 0.63 (0.33) & 0.08 (0.11)  \\
  $x + \text{CW}$       & 0.49 (0.31) & 0.84 (0.26) & 0.49 (0.26) & 0.28 (0.25) & 0.85 (0.36) & 0.10 (0.14)  \\
  \hline
\end{tabular}
\caption{Mean (Standard Deviation) WER of transcriptions for our experiments} 
\label{tab:werresults}
\end{table*}

\section{Discussion}
Upon investigation, SOTA STT systems fail to transcribe our attacks accurately, which can be represented in terms of the attack audio's similarity with the original, and audio signal properties.

\subsubsection{Cosine Similarity Analysis}\label{sec:cosinesimilarity}
We posit that our attack can successfully fool STT systems due to the similarity of the attack to its original audio. We note that the mean normalised cosine similarity of our attacked audio and the original is very close to 1.0, surpassing the CW attacked audio. Table \ref{tab:cosinesimilarity} shows the similarities of our attacks.

\begin{table}[ht]
\centering
\begin{tabular}{p{2cm} p{2cm} p{2.5cm}}
  \hline
  Audio File & Mean & Standard Deviation \\  
  \hline
  $x + \delta_0$        & 0.999930 & 0.0000337\\
  $x + \delta_{-5}$     & 0.999962 & 0.0000216\\
  $x + \delta_{-10}$    & 0.999979 & 0.0000142\\
  $x + \delta_{-15}$    & 0.999989 & 0.0000095\\
  $x + \delta_{-20}$    & \textbf{0.999994} & \textbf{0.0000063}\\
  $x + \text{CW}$       & 0.999961 & 0.0000229\\
  \hline
\end{tabular}
\caption{Cosine similarities between the original audio and the adversarial audio} 
\label{tab:cosinesimilarity}
\end{table}

\subsubsection{Audio Signal Analysis}\label{sec:spectrogramanalysis}
We analyse the original signal and our adversarial audio signal using audio signal analysis methods: Fast Fourier Transform (FFT) and Spectrograms.

\begin{figure*}[h]
  \centering
  \begin{subfigure}[b]{0.40\textwidth}
    \includegraphics[width=\linewidth]{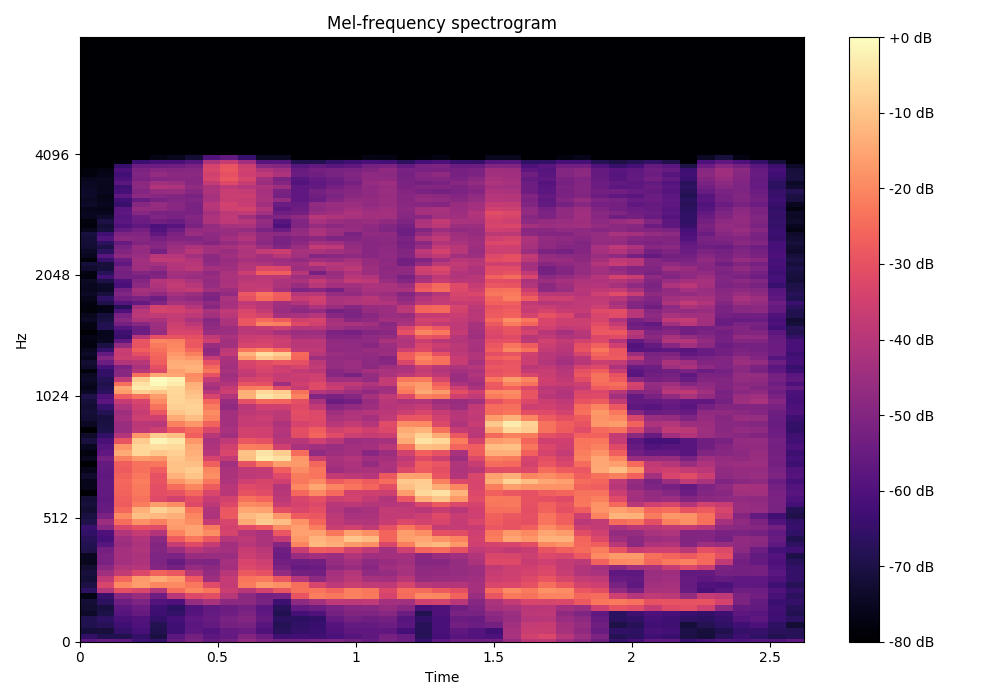}
  \end{subfigure}
  \begin{subfigure}[b]{0.40\textwidth}
    \includegraphics[width=\linewidth]{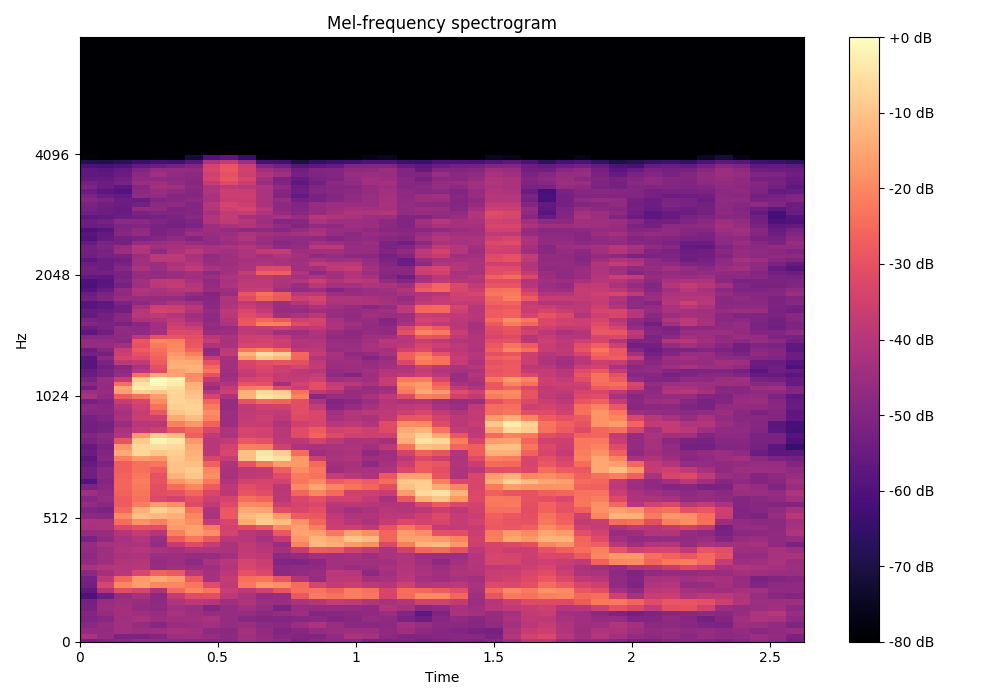}
  \end{subfigure}
  \begin{subfigure}[b]{0.40\textwidth}
    \includegraphics[width=\linewidth]{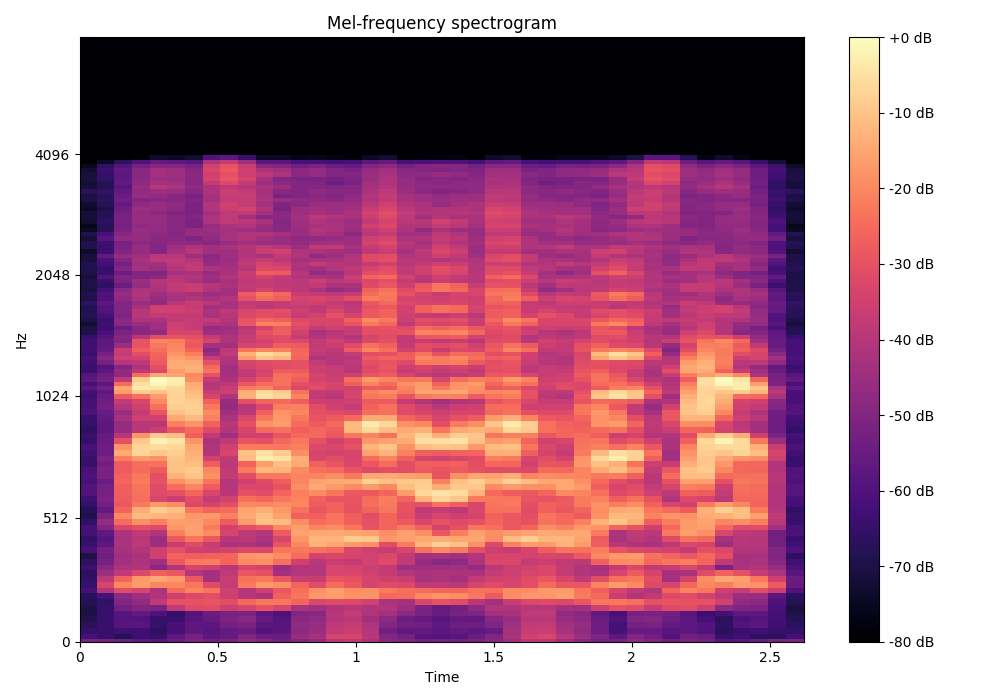}
  \end{subfigure}
  \begin{subfigure}[b]{0.40\textwidth}
    \includegraphics[width=\linewidth]{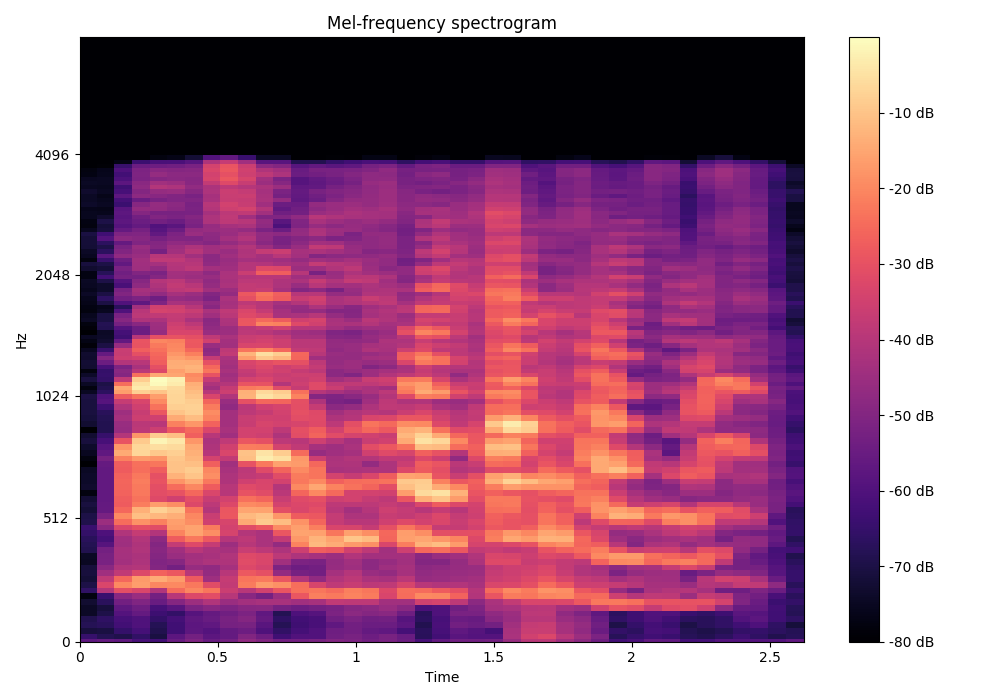}
  \end{subfigure}
  \begin{subfigure}[b]{0.40\textwidth}
    \includegraphics[width=\linewidth]{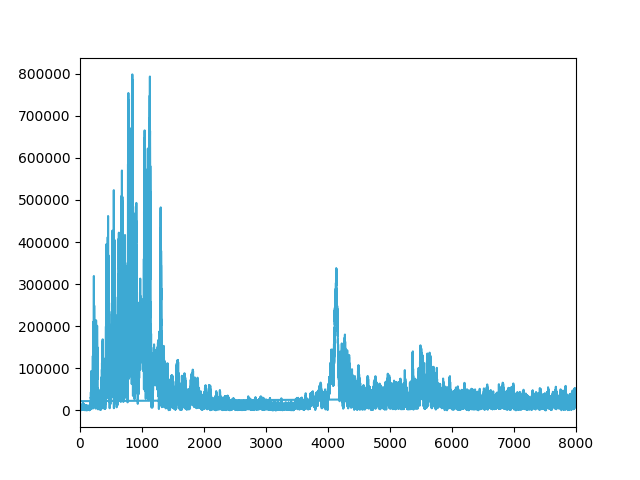}
     \caption*{Original audio $x$}
  \end{subfigure}
  \begin{subfigure}[b]{0.40\textwidth}
    \includegraphics[width=\linewidth]{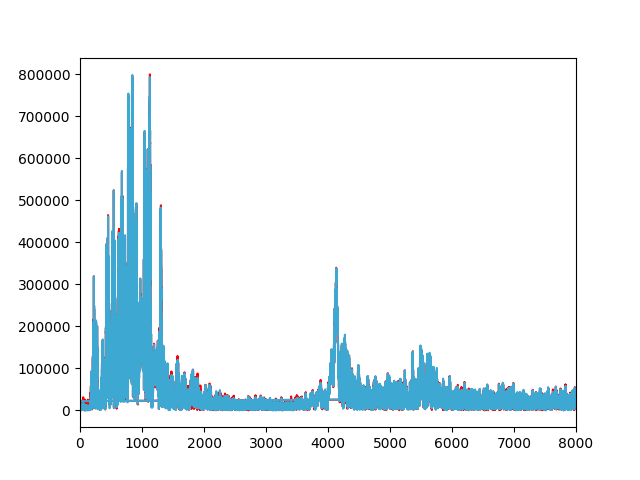}
    \caption*{$x + \text{CW}$}
  \end{subfigure}
  \begin{subfigure}[b]{0.40\textwidth}
    \includegraphics[width=\linewidth]{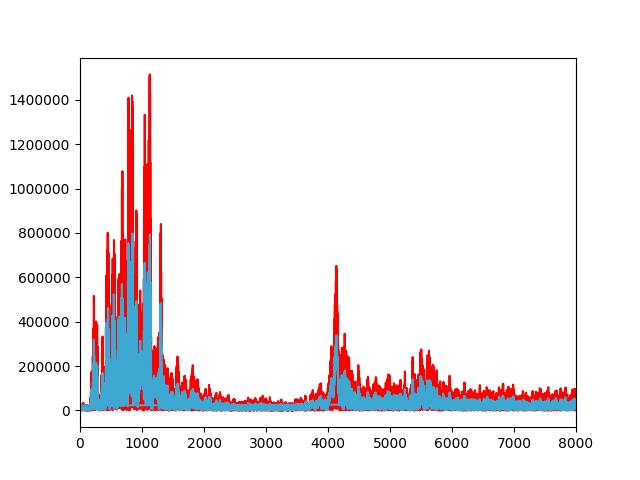}
    \caption*{$x + \delta_0$ }
  \end{subfigure}
  \begin{subfigure}[b]{0.40\textwidth}
    \includegraphics[width=\linewidth]{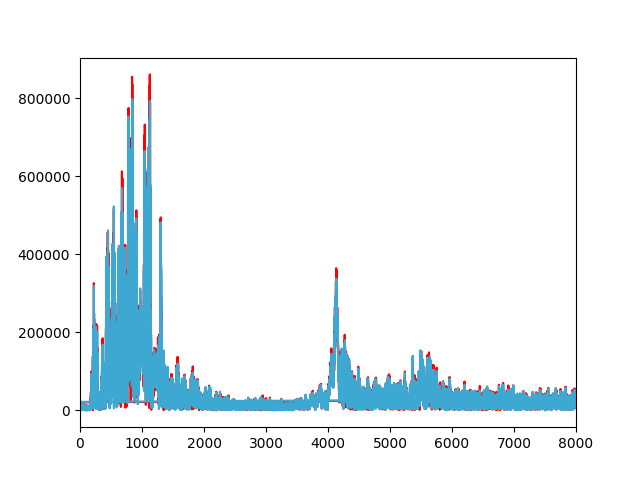}
    \caption*{$x + \delta_{-15}$}
  \end{subfigure}
  \caption{Mel-frequency spectrograms and FFT overlays of audio samples. We observe that the original vocal masks are preserved in the CW attacks, but our attacks generate additional vocal masks that can effectively fool vocal mask analysis methods.}
  \label{fig:spectrograms}
\end{figure*}

Figure \ref{fig:spectrograms} presents the spectrograms and FFT plots of our audio adversarial examples. We note that from the spectrograms generated from $x + \text{CW}$, the CW attack possess clear vocal masks of the original audio, which makes it possible to retrieve some semblance of the original audio using a vocal mask technique, built upon computer vision principles. Our attacks obfuscate the original audio by appearing as the original audio, yet in the opposite direction, hence are able to fool techniques that employ vocal mask analyses. In the case of the FFT plots, the plot generated from $x + \text{CW}$ is hugely similar to the plot of $x$, which means it is possible to retrieve $x$ from the adversarial audio. However, our adversarial audio has different FFT forms compared to $x$, since our attack introduces additional audio signals, thereby preventing attempts to recover the original audio.

\subsubsection{Perception of Speech}
Our results reflect that humans are better at the transcription task, regardless of obfuscation. Humans have innate experience with language and are able to predict words in a sentence even if they did not hear it from the audio. This phenomenon is known as auditory perceptual restoration, where the brain fills in missing information in areas where noise obstructs portions of sounds \cite{BIDELMAN201684}\cite{auditoryillusion}. Commonly observed in a conversation with loud background noise, this phonemic restoration effect occurs where the brain restores sounds missing from a speech signal. This effect is typically observed when missing phonemes in an auditory signal are replace with noises that masks the original phonemes, creating an auditory ambiguity\cite{perceptual}\cite{brain}. In the case of our experiments, our attack audio masks some original audio, which causes perceptual restoration to kick in, where the humans perform better transcriptions due to this phenomenon.

\subsubsection{Future Work}
From this work, we acknowledge that humans possess an innate language ability which allowed them to perform better in the transcription task. We posit that future STT systems can include predictive abilities, such as introducing generative models with attention gates that govern which audio features have more importance, thereby allowing STT systems to predict transcriptions and at the same time inference them directly from the audio files.

\section{Conclusion}
We demonstrate a novel method of creating audio adversarial examples by reversing the audio signal and overlaying it on the original speech. We present evidence that these adversarial examples render vocal masks obsolete due to the inability to identify the reversed audio from the true audio. Our experiments show that these adversarial examples fool State-Of-The-Art Speech-To-Text systems, yet humans are able to consistently pick out the speech.  We hope that future work will continue to investigate audio adversarial examples, and improve STT systems with predictive language abilities that humans possess.

\bibliographystyle{acl}
\bibliography{bibliography.bib}

\end{document}